\documentclass[aps,prl,twocolumn,groupedaddress,showpacs]{revtex4}
\usepackage{graphicx}

\begin{document}

\title{High-pressure phases of silane}

\author{Chris J. Pickard}
\author{R. J. Needs}
\affiliation{Theory of Condensed Matter Group, Cavendish Laboratory,
J.~J.~Thomson Avenue, Cambridge CB3 0HE, United Kingdom}

\date{\today}

\begin{abstract}

It has been suggested that hydrogen may metallise at lower pressures
if it is ``precompressed''. Here we introduce a search strategy for
predicting high-pressure structures and apply it to silane using
first-principles electronic structure computations. It is based on
relaxing randomly chosen structures, and is demonstrated to work well
for unit cells containing up to at least ten atoms. We predict that
silane will metallise at higher pressures than previously anticipated,
but we suggest that the metallic phase might show high-temperature
superconductivity at experimentally accessible pressures.

\end{abstract}


\maketitle

As early as 1935 Wigner and Huntingdon suggested that hydrogen would
become metallic under sufficient compression.\cite{WignerH35} There
have been many attempts to produce metallic hydrogen under
high-pressure laboratory conditions, but it has stubbornly remained
insulating in diamond-anvil-cell experiments up to the highest
pressure so far achieved of 342~GPa.\cite{NarayanaLOR98} It has
recently been suggested\cite{Ashcroft04} that the group IV hydrides,
methane (CH$_4$), silane (SiH$_4$), and germane (GeH$_4$), might
become metallic at pressures achievable in diamond anvil cells because
the hydrogen in these materials is ``chemically precompressed'' by the
presence of the group IV atoms.  Such metallic phases might exhibit
phonon-mediated high-temperature superconductivity.

Recently, Feng \textit{et al.}\cite{FengGJHBA06} reported a
first-principles density-functional-theory (DFT) study of silane at
high pressures.  They predicted that a layered structure would be
stable at pressures above 25~GPa.  This phase was found to be
semi-metallic at 91~GPa, and the density of electronic states at the
Fermi energy increased rapidly with further application of pressure.
An estimate using the Bardeen-Cooper-Schrieffer (BCS)
theory~\cite{BardeenCS57} indicated that the superconducting
transition temperature would increase from about 0~K at 91~GPa to
about 166~K at 202~GPa, pressures which are comfortably within range
of diamond anvil cells.

Accurate cohesive energies and equilibrium volumes for materials of
known crystal structure have been obtained using first-principles
methods, but the prediction of stable structures has remained very
difficult.  Using these methods, it is routine to relax a structure
from a chosen initial configuration to a nearby minimum in the
potential energy surface.  Relaxing from a set of initial
configurations may lead to several distinct local-energy minima, the
lowest-energy one corresponding to the most favoured structure.
However, if none of the initial configurations is sufficiently close
to the global minimum then the configurations will become trapped in
higher-energy local minima.  Trapping in local-energy minima is a
severe problem because the number of minima is expected to increase
exponentially with the number of degrees of freedom.  
In the past this problem has typically been addressed by selecting
initial configurations derived from experimental information known
about the system in question, the known structures of similar
materials, chemical intuition, and results from simpler computational
methods. The shortcomings in this approach have led to a growth in using
first-principles methods in combination with more advanced search
techniques to determine stable
structures.\cite{MartonakLP03,YooZ05,GoedeckerHL05,OganovGO06}

We have developed a simple strategy for generating initial
configurations, which consists of choosing essentially random unit
cell parameters and atomic positions.  We first choose the number and
identity of the atoms in the unit cell.  To form ``random'' unit cells
we choose three cell-vector lengths randomly and uniformly between 0.5
and 1.5 (in arbitrary units), and three cell angles randomly and
uniformly between 40$^{\circ}$ and 140$^{\circ}$.  The resulting cell
vectors are scaled to produce the desired volume, which is also
selected randomly and uniformly from between 0.5 and 1.5 of some
chosen volume.  The positions of the atoms are then chosen randomly
and uniformly within the cell.  First-principles methods are then used
to relax each structure until a minimum in the enthalpy is reached.
The structures obtained from such a procedure are clearly liable to be
trapped in local minima, and undoubtedly the probability of finding
the global minimum will drop rapidly with increasing complexity of the
system.  However, for systems of moderate complexity, tests indicate
that our approach has a high probability of locating the global
minimum.  In practice, the number of initial configurations required
to find the global minimum depends on the system, and particularly on
the number of degrees of freedom it possesses.  Our approach has been
to continue generating configurations until the relaxed structures
with low energies are generated several times and, where possible, to
look for the occurrence of previously-known ``marker'' structures.

All of the first-principles calculations were performed using a
developer's version of the CASTEP code.\cite{SegallLPPHCP02} For the
initial search over structures we used the local density approximation
(LDA) for the exchange-correlation functional\cite{PerdewZ81}, and
default ultrasoft pseudopotentials.\cite{Vanderbilt90} The $k$-point
sets were generated separately for each unit cell encountered during
the procedure, and no symmetry restrictions were applied at any point.
Medium quality Brillouin zone sampling using a grid of spacing $2\pi
\times$~0.07~\AA$^{-1}$ and a plane wave basis set cutoff of 280~eV
were found to be sufficient for the initial search over structures.
When re-calculating the enthalpy curves with higher accuracy we used a
Brillouin zone sampling of $2\pi \times$~0.03~\AA$^{-1}$, harder
pseudopotentials requiring a plane wave cutoff of 360~eV, and the
Perdew-Burke-Ernzerhof (PBE) Generalised Gradient Approximation (GGA)
density functional\cite{PerdewBE96} to aid comparison with
Ref.~\cite{FengGJHBA06}.  Each of the calculations was for the
zero-temperature ground state and we neglected the zero-point
vibrational energy.

We have applied our approach to silicon at zero pressure.  Using
two-atom unit cells and 100 randomly chosen initial configurations, we
obtained the global minimum energy diamond structure (15 times),
$\beta$-tin (2 times), Imma (14 times), and simple hexagonal (15
times).  Both experiment and DFT calculations agree that these phases
are the four lowest-pressure thermodynamically stable phases of
silicon.\cite{MujicaRMN03} A study of 8-atom unit cells of carbon at
300~GPa generated the diamond structure as the lowest-enthalpy phase
in addition to numerous higher-enthalpy phases, including the BC8
phase, which has been postulated as a stable phase of carbon at TPa
pressures.\cite{Yin84} We also studied 16-atom unit cells of hydrogen
at 250~GPa, finding, among other phases, the $Cmca$ molecular phase
which DFT calculations have predicted to be the most stable at that
pressure.\cite{JohnsonA00}

Our approach is particularly suitable for the prediction of the
structures of high-pressure phases because they often have fairly
simple structures with a small number of atoms in the primitive unit
cell. In addition, our method is straightforward and does not require
the selection of highly-system-specific parameter values. Our scheme
could be modified to include other features, such as methods for
extricating configurations from local energy minima.\cite{Wales03}
However, such modifications would complicate the method and add to the
computational cost, which would limit the number of initial
configurations which can be sampled.  It is also possible to introduce
other constraints on the initial configurations so that one can
predict the structures of clusters, defects, surfaces, etc.  For
example, we have used our approach to search for low-energy
self-interstitial defect structures in silicon.  To construct the
initial configurations we took a 32-atom body-centred-cubic unit cell
of diamond-structure silicon, removed an atom and its four nearest
neighbours, and then placed six silicon atoms randomly within a
spherical region of radius about 1.5 bond lengths centred on the first
atom that we removed.  Using our approach, we readily found the
split-$\langle110\rangle$ and hexagonal interstitial configurations,
which other DFT calculations have shown to be lowest in
energy\cite{Needs99}, and also predicted that all other
self-interstitial local energy minima are at least 0.3~eV higher in
energy.  The generation of numerous low-energy structures is an
interesting and useful feature of our strategy, as they can give
information about the structures which might be formed under other
conditions of pressure and temperature, or out of thermal equilibrium.

We tested our search strategy for simulation cells containing one,
two, and four SiH$_4$ units at a pressure of 250~GPa.  With one
SiH$_4$ unit per cell the lowest-enthalpy structure occurred four
times from 13 initial configurations, but it was 0.36~eV per SiH$_4$
unit higher in enthalpy than the most stable structure found with two
SiH$_4$ units per cell, which occurred 8 times from 37 initial
configurations.  With four SiH$_4$ units per cell the most stable
structure was 0.13~eV per SiH$_4$ unit higher in enthalpy than for the
two SiH$_4$ unit cell, and each of the low-enthalpy phases occurred
only once from 38 configurations.  Clearly the simulation cells
containing one SiH$_4$ unit are too constrained to produce
very-low-enthalpy structures, while those containing four SiH$_4$
units have too many degrees of freedom for our search strategy to find
the global minimum enthalpy structure with the number of initial
configurations used.  We concluded that using two SiH$_4$ units per
simulation cell and 40 initial configurations represented an
affordable and efficient approach to finding low-enthalpy phases of
silane at high pressures.

Further runs were performed at 0, 50, 100, 150, and 200~GPa, with two
SiH$_4$ units per simulation cell and 40 initial structures.  At 0~GPa
the lowest-energy structure could not be determined with confidence
because the low-energy structures were not found repeatedly.  This
reflects the fact that at zero pressure many different structures have
similar energies (within 0.1~eV per SiH$_4$ unit), including those
consisting of silane molecules, disilane plus a hydrogen molecule, and
various polymeric forms plus hydrogen molecules.  The various packings
of silane molecules were therefore inadequately explored using just 40
initial structures.  The lowest-energy structure found (by about
40~meV per SiH$_4$ unit) was a low-symmetry silane molecular crystal
which was about 50~meV per SiH$_4$ unit less stable than the
higher-symmetry T1 packing considered by Feng \textit{et
al.}~\cite{FengGJHBA06}.

\begin{figure}
\includegraphics[width=0.5\textwidth]{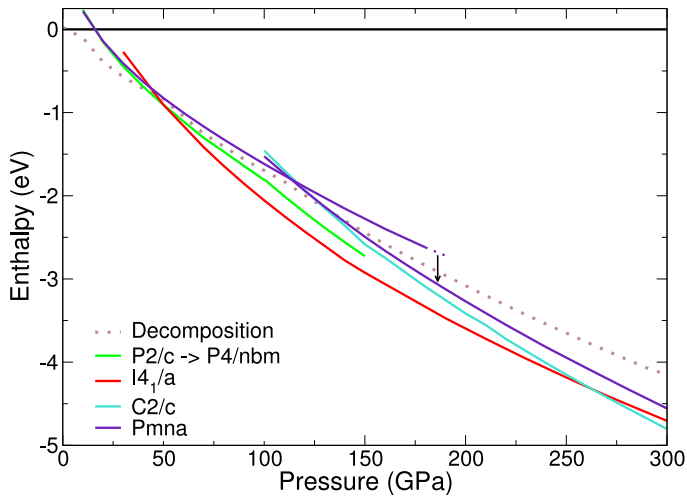}
\caption{\label{fig:enthalpy} The enthalpies per SiH$_4$ unit of
various structures as a function of pressure, referenced to the T1
phase of Ref.~\cite{FengGJHBA06}.}
\end{figure}

\begin{figure}
\includegraphics[width=0.45\textwidth]{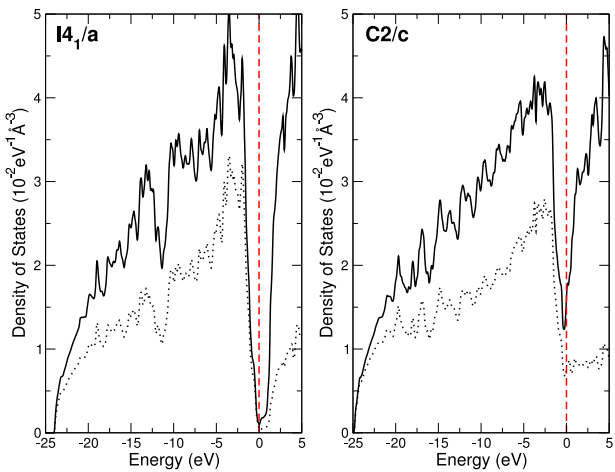}
\caption{\label{fig:dos} The densities of states of the $I4_1/a$ and
$C2/c$ phases at their predicted coexistence pressure of
262.5~GPa. The partial density of states projected onto the H atoms is
shown by dashed curves, and the Fermi energies are indicated by the
vertical dashed lines.}
\end{figure}

Having selected the low-enthalpy phases from the above runs, we
re-calculated their enthalpies with greater computational accuracy.
The calculated enthalpies of various structures are shown as a
function of pressure in Fig.~\ref{fig:enthalpy}.  The details of
some of the lower-enthalpy structures found are described in
the auxiliary material to this Letter.\cite{EPAPS}

An insulating chain-like structure with $P2/c$ symmetry was found to
be the most stable in a small region of pressures around 40~GPa, but a
structure of $I4_1/a$ symmetry became more stable at about 50~GPa.
The $I4_1/a$ phase is predicted to be insulating at 50~GPa but, as the
pressure is increased, the bandgap gradually closes and it becomes
semi-metallic, as can be seen in the density of states plotted in
Fig.~\ref{fig:dos}.  Given the well-known shortcomings of standard DFT
approaches such as the LDA and GGA functionals, which normally
underestimate band gaps, it is likely that our calculations
underestimate the pressure at which the $I4_1/a$ phase becomes
semi-metallic.  The $I4_1/a$ phase remains the lowest-enthalpy phase
up to 262.5~GPa, at which pressure a denser structure of $C2/c$
symmetry becomes stable.  This phase is a good metal, see
Fig.~\ref{fig:dos}.

\begin{figure}
\includegraphics[width=0.25\textwidth]{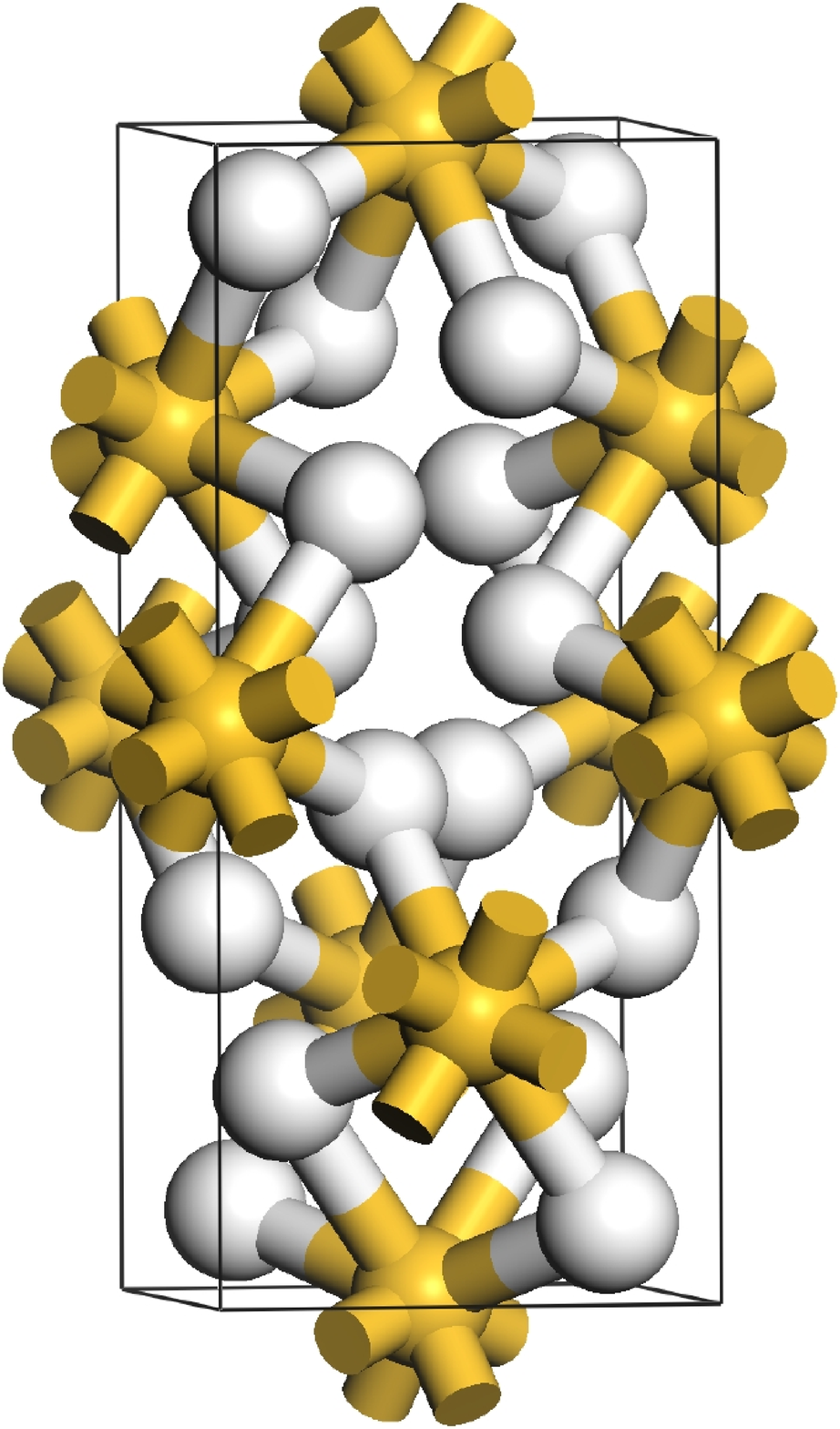}\includegraphics[width=0.25\textwidth]{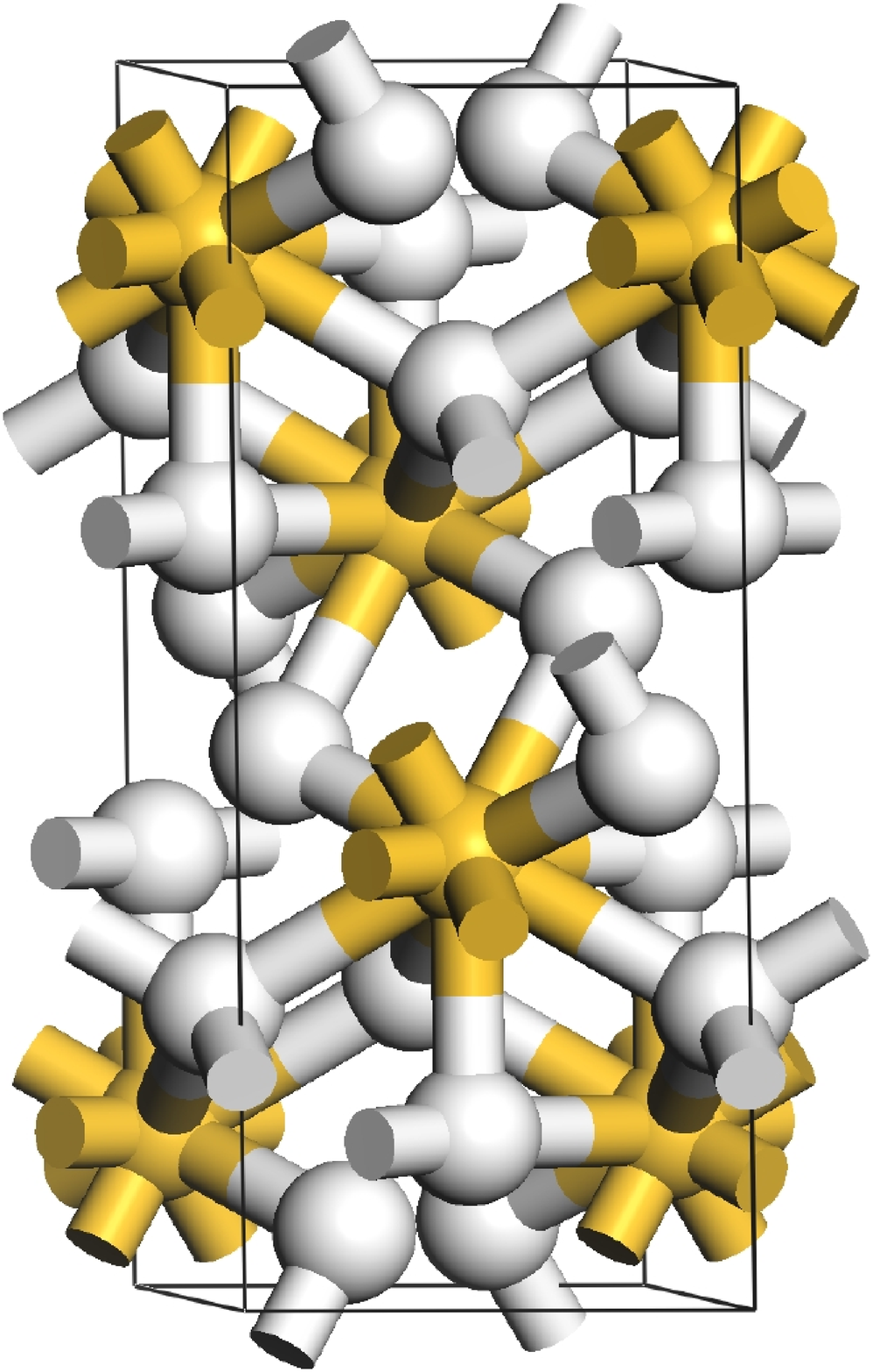}\\
\centerline{\large\textsf{I4}$_{\sf1}/$\textsf{a}\hspace{3.0cm}\textsf{C2/c}\normalsize}
\caption{\label{fig:structures} The $I4_1/a$ and $C2/c$ structures.
The golden spheres represent silicon atoms, and the white spheres
hydrogen. Note the Si$_2$H$_2$ planes in the $I4_1/a$ structure, and
the two, three and four-fold coordinated H atoms in the $C2/c$
structure.}
\end{figure}

In the $I4_1/a$ structure, which is illustrated in
Fig.~\ref{fig:structures}, each Si atom is bonded to 8 H atoms.  Each
of the H atoms forms a ``bridge'' between two neighbouring Si atoms.
There are two such bridges between every pair of neighbouring Si
atoms, the four atoms forming a Si$_2$H$_2$ plane.  This bonding
arrangement consists of two ``banana bonds'', which are
electron-deficient three-centre-two-electron bonds, reminiscent of
those in B$_2$H$_6$ (diborane).\cite{BurnelleK65}
The arrangement of the Si$_2$H$_2$ planes keeps the H atoms away from
each other, and the $I4_1/a$ structure contains only Si-H bonds.  Each
of the Si sites and the H sites is equivalent in this high-symmetry
structure which has a 10-atom primitive unit cell.  The layered O3
structure proposed in Ref.~\cite{FengGJHBA06} also contains some
bridging Si-H-Si bonds, although it also contains H atoms which are
bonded to single Si atoms.  Our $I4_1/a$ structure is about 0.4~eV per
SiH$_4$ unit lower in enthalpy than the O3 structure at 100~GPa, and
about 0.7~eV per SiH$_4$ unit lower at 180~GPa.

In the metallic $C2/c$ structure illustrated in
Fig.~\ref{fig:structures}, each Si atom is bonded to 11 H atoms.  Each
of the Si sites is equivalent and there are three inequivalent H sites
in which the H atom forms bonds to two, three, or four Si atoms.  The
average coordination number of the H atoms in the $C2/c$ structure is
larger than in $I4_1/a$, which accounts for its smaller equilibrium
volume.  The metallic character of this phase means that the nature of
the inter-atomic bonding is less clear cut than in the $I4_1/a$
structure, but again a description in terms of electron deficient Si-H
bonds is appropriate.

Our search also revealed a number of other low-enthalpy phases,
including the O3 $Pmna$ structure studied by Feng \textit{et
al.}\cite{FengGJHBA06} We found that, at around 200~GPa, O3 $Pmna$
converts into a more compressed structure of the same symmetry, but it
still has a substantially higher enthalpy than our $I4_1/a$ phase.  We
followed the more compressed phase to lower pressures and found it to
be more stable than the O3 $Pmna$ structure down to about 110~GPa.  We
also found a low-enthalpy phase with $I\overline{4}2d$ symmetry which
is only 0.1~eV per SiH$_4$ unit above the most stable phase at
100~GPa.  The $P2/c$ phase mentioned above converts to a structure
with $P4/nbm$ symmetry at 100~GPa. Interestingly, the $P4/nbm$ phase
has a layered structure, like the O3 structure, but with every
hydrogen atom participating in a bridging bond, which results in its
lower enthalpy.

Having found the low-enthalpy metallic $C2/c$ structure using
simulation cells containing two SiH$_4$ units, we studied its
stability to rearrangements of the H atoms using simulation cells
containing four SiH$_4$ units at 250~GPa.  To make initial
configurations we used the $C2/c$ Si framework and placed the 16 H
atoms randomly within the simulation cell, and then relaxed all of the
atomic positions and cell parameters.  Using 15 such initial
configurations we obtained the $C2/c$ structure as the lowest-enthalpy
phase, an $Fddd$ structure only 30~meV per SiH$_4$ unit higher in
enthalpy, and several other slightly less stable phases.  All of the
low-enthalpy phases contained a mixture of two-, three-, and four-fold
coordinated H atoms, as in the $C2/c$ structure, and they all had very
similar volumes and densities of electronic states.

Silane is in fact unstable to decomposition into hydrogen and silicon
at zero pressure, and we therefore considered the thermodynamic
stability of our new silane phases against processes involving the
formation of hydrogen molecules.  At low pressures we found some
low-enthalpy structures containing covalently bonded hydrogen
molecules, indicating the instability of silane at these pressures,
but none of the low-enthalpy structures at high pressures contain any
H-H bonds, suggesting stability against decomposition.  We
investigated this further by considering decomposition into
diamond-structure Si and molecular hydrogen at zero pressure, into
simple hexagonal silicon\cite{MujicaRMN03} and $Cmca$-structure
hydrogen\cite{JohnsonA00} phases up to 40~GPa, and into face-centred
cubic silicon and $Cmca$-structure hydrogen at higher pressures. While
the O3 structure proposed in Ref.~\cite{FengGJHBA06} is found to be
unstable with respect to decomposition into face-centred cubic silicon
and $Cmca$-structure hydrogen at its metallization pressure, our
$I4_1/a$ and $C2/c$ structures are stable with respect to such a
decomposition.  This lends further weight to our contention that the
$C2/c$ structure of silane, or something very like it, is
thermodynamically stable at high pressures.

We now discuss the possibility of phonon-mediated superconductivity in
the $C2/c$ phase.  Within the BCS theory, the superconducting
transition temperature depends on the values of three parameters: the
density of states at the Fermi energy of the normal state, the Debye
temperature of the phonons, and the effective electron-phonon coupling
parameter.  We estimated the densities of states at the Fermi energies
of the $C2/c$ and $I4_1/a$ phases at their coexistence pressure of
262.5~GPa using DFT methods. As shown in Fig.~\ref{fig:dos}, the
$C2/c$ phase is a good metal, with a density of states at the Fermi
energy of $1.6 \times 10^{-2}$~eV$^{-1}$ \AA, which is about 46\% of
the free-electron value and is close to the zero-pressure value for
lead.
We decomposed the densities of states into H- and Si-associated
components using a Mulliken population analysis.\cite{SegallSPP96}
Fig.~\ref{fig:dos} shows that below the Fermi energy the density of
states is a little larger on the H atoms than on the Si ones, while
above the Fermi energy it becomes progressively larger on the Si
atoms.  This is consistent with the fact that the Pauling
electronegativity of H (2.2) is a little larger than that of Si (1.9).
%Pauling electronegativities: H: 2.2, C: 2.55, Si, 1.90, Ge, 2.01.
The fact that the density of electronic states at the Fermi energy
projected onto the H atoms is large suggests that the coupling between
the electrons close to the Fermi energy and phonon modes involving
motions of the H atoms may be significant. We also performed DFT
calculations of the $\Gamma$-point phonon frequencies of the $C2/c$
phase at 262.5~GPa, finding the highest phonon frequency to be
$2500$~cm$^{-1}$. This corresponds to an estimated Debye temperature
of 3600~K, which is about 40 times larger than that of lead at
zero-pressure.  An examination of the phonon eigenvectors indicates
that the highest-frequency modes correspond largely to motions of the
H atoms.  The very large Debye temperature and the substantial density
of states at the Fermi energy suggests that the $C2/c$ phase might
exhibit high-temperature superconductivity, but one should be cautious
about this conclusion.  A reliable estimate of the superconducting
transition temperature within the BCS theory also requires evaluation
of the full electron-phonon coupling matrix elements, which is beyond
the scope of this work.

In summary, we have used a random searching strategy in conjunction
with first-principles electronic structure computations to predict the
stable high-pressure phases of silane.  We find
insulating/semi-metallic behaviour up to 260~GPa, at which point a
first-order phase transition occurs to a good metal.  This phase might
show high-temperature superconductivity at a pressure which is
achievable within a diamond anvil cell.

\begin{table*}
\begin{ruledtabular}
\begin{tabular}{cclllllll}
Pressure       & Space group     & \multicolumn{3}{c}{Lattice parameters}           & \multicolumn{4}{c}{Atomic coordinates} \\
(GPa)          &                 & \multicolumn{3}{c}{(\AA, $^{\circ}$)}            & \multicolumn{4}{c}{(fractional)}       \\\hline
50             & $P2/c$          & $a$=4.70       & $b$=4.02       & $c$=3.48       & Si & 0.5000 & 0.2078 & 0.7500          \\
               &                 & $\alpha$=90.00 & $\beta$=137.34 & $\gamma$=90.00 & H1 & 0.7745 & 0.1272 & 0.6642          \\
               &                 &                &                &                & H2 & 0.1919 & 0.4024 & 0.2277          \\\\
100            &$I\overline{4}2d$& $a$=4.26       & $b$=4.26       & $c$=3.95       & Si & 0.0000 & 0.0000 & 0.0000          \\
               &                 & $\alpha$=90.00 & $\beta$=90.00  & $\gamma$=90.00 & H  & 0.3285 & 0.0367 & 0.1195          \\\\
150            &$P4/nbm$         & $a$=3.35       & $b$=3.35       & $c$=2.80       & Si & 0.5000 & 0.5000 & 0.5000          \\
               &                 & $\alpha$=90.00 & $\beta$=90.00  & $\gamma$=90.00 & H  & 0.3553 & 0.0367 & 0.1195          \\\\
150            &$I4_1/a$         & $a$=3.04       & $b$=3.04       & $c$=6.85       & Si & 0.5000 & 0.5000 & 0.5000          \\
               &                 & $\alpha$=90.00 & $\beta$=90.00  & $\gamma$=90.00 & H  & 0.8676 & 0.2166 & 0.4328          \\\\
250            &$C2/c$           & $a$=2.94       & $b$=6.79       & $c$=2.90       & Si & 0.5000 & 0.6334 & 0.2500          \\
               &                 & $\alpha$=90.00 & $\beta$=115.20 & $\gamma$=90.00 & H1 & 0.8028 & 0.4462 & 0.4986          \\
               &                 &                &                &                & H2 & 0.0000 & 0.6389 & 0.7500          \\
               &                 &                &                &                & H3 & 0.0000 & 0.7549 & 0.2500        
\end{tabular}
\end{ruledtabular}
\caption{\label{table:structures} {{ Details of some of the
structures produced by the search.} The structures have been optimised
at the higher level of accuracy described in the letter. Only
the fractional coordinates of symmetry inequivalent atoms are
reported.}}
\end{table*}

\begin{acknowledgments}
CJP was supported by an EPSRC Advanced Research Fellowship.  We thank
Peter Littlewood for useful discussions, and Keith Refson for
assistance in performing the phonon calculations.
\end{acknowledgments}

\end{document}